\documentclass[pra,twocolumn,superscriptaddress,superscriptaddress,amsmath]{revtex4-2}
 \usepackage{graphicx}
\usepackage{float}
\usepackage{epstopdf,cancel}
\usepackage{epsf,latexsym,bbm}
\usepackage[mathscr]{euscript}

\usepackage{amssymb,amsmath}
\usepackage{mathtools} 
\usepackage{times,graphics}
\usepackage{soul,xcolor}
 \usepackage{epsfig,ulem}
\def\6{\langle}
\def\9{\rangle}

\newcommand\SC{{\mathrm{{sc}}}}
\newcommand\GS{{\mathrm{gs}}}
\def\sc{{\mathrm{sc}}}
\newcommand\gs{{\mathrm{gs}}}

\def\arb{\mathsf{b}}

\def\af{\mathsf{f}}
\def\ak{{\mathsf{k}}}

\def\bob{\mathbf{b}}

\def\bof{\mathbf{f}}
\def\bg{\mathbf{g}}
\newcommand\bk{{\bf{k}}}
\def\bv{\mathbf{v}}
\def\bu{\mathbf{u}}

\def\bE{\mathbf{E}}

\newcommand\sg{\textsl{g}}

\newcommand\co{{\cal{O}}}
\newcommand\eR{{\mathscr{R}}}

\def\hbb{{\hat{\bob}}}

\def\hbf{{\hat{\bof}}}
\def\hbk{{\hat{\bk}}}
\def\hbv{\hat{\mathbf{v}}}

\def\hbx{\hat{\mathbf{x}}}
\def\hby{\hat{\mathbf{y}}}
\def\hbz{\hat{\mathbf{z}}}

\newcommand\vb{{\vec{b}}}
\newcommand\vd{{\vec{d}}}
\newcommand\ve{{\vec{e}}}
\newcommand\vf{{\vec{f}}}
\newcommand\vk{{\vec{k}}}
\newcommand\vl{{\vec{l}}}
\newcommand\vn{{\vec{n}}}
\newcommand\vv{{\vec{v}}}
\newcommand\vx{{\vec{x}}}
\newcommand\vJ{{\vec{J}}}

\def\vkap{{\vec{\kappa}}}

\newcommand{\Omegab}{\mbox{\boldmath$\Omega$}}
\newcommand{\omegab}{\mbox{\boldmath$\omega$}}

\newcommand{\hzeb}{\mbox{\boldmath$\hat\zeta$}}

\def\half{\tfrac{1}{2}}

\newcommand{\pad}{\partial}
\newcommand{\be}{\begin{equation}}
\newcommand{\ee}{\end{equation}}
\newcommand{\ba}{\begin{eqnarray}}
\newcommand{\ea}{\end{eqnarray}}

\newcommand{\etal}{\textit{et al.}}

\newcommand{\defeq}{\vcentcolon=}
\newcommand{\eqdef}{=\vcentcolon}

\usepackage{url,hyperref}
\hypersetup{colorlinks,linkcolor={blue!55!black},citecolor={red!45!black},urlcolor={blue!45!black},breaklinks=true}

\begin{document}

\title{Polarization rotation and near-Earth quantum communications}
\author{Pravin Kumar Dahal}
     \affiliation{Department of Physics \& Astronomy,
 Macquarie University, Sydney NSW 2109, Australia}

\author{Daniel R. Terno}
\affiliation{Department of Physics \& Astronomy, Macquarie University, Sydney NSW 2109, Australia}

\begin{abstract}
       We revisit polarization rotation due to gravity, known as the gravitational Faraday effect, with a view on its role in quantum communications with   Earth-orbiting satellites. In a static spherically symmetric  gravitational field Faraday rotation is purely a reference frame (gauge) effect. This is so also in the leading post-Newtonian expansion of the Earth's gravitational field. However, establishing  the local reference frame with respect to distant stars leads to the nonzero Faraday phase. In communications  between a ground station and an Earth-orbiting spacecraft this phase is of the order of $10^{-10}$. Under the same conditions the Wigner phase of special relativity is typically of the order $10^{-4}$--$10^{-5}$. These phases lead to the physical lower bound on communication errors. However, both types of errors can be simultaneously mitigated. Moreover, they are countered by a fully reference frame independent scheme that also handles arbitrary misalignments between the reference frames of sender and receiver. 

\end{abstract}
\maketitle

\section{Introduction}
Space deployment of quantum technology \cite{micius-1,time,qsat} brings it into
a weakly relativistic regime. As a side effect,  low-Earth  orbit (LEO) quantum communication satellites   provide  new opportunities to test  fundamental physics. Once these tiny  physical effects fall within the
sensitivity range of these devices, they may impose constraints on practical quantum
communications, time-keeping, or remote sensing tasks. This dual relationship between quantum technology and relativistic physics  makes it important to study the relativistic aspects of quantum information processing \cite{pt:04,cqg,qs:12}.

 Qubits are routinely realized as  polarization states of photons \cite{photon}. The effects of special relativity (SR) on quantum-informational tasks with massive and massless particles are well-understood \cite{pt:04,qs:12}. On the other hand, effects of general relativity (GR) in quantum-optical LEO experiments were discussed primarily in the context of interferometry \cite{cqg,bf:14,inter:15,red}.

 The dominant  source of relativistic errors in this setting is   the Wigner rotation (or phase), a SR effect \cite{pt:04,qs:12}.  Gravitational polarization rotation, also known as the gravitational Faraday effect \cite{gfaraday-1,gfaraday-2} occurs in a variety of astrophysical systems, such as accretion disks around astrophysical black hole candidates \cite{f-bh} or gravitational lensing \cite{f-gl}. This effect was the subject  of a large number of theoretical investigations, first in the framework of geometric optics \cite{gfaraday-1,gfaraday-2,f-gl,fl:82,tg-pol}, and  recently in the post-eikonal approximation \cite{post-f,post-f2}.  Interpretation of these results is sometimes contradictory, as the crucial role of local reference frames and the ensuing introduction of the standard polarization directions   were not always treated   in a fully satisfactory manner.

The Faraday effect in quantum experiments was discussed in \cite{fara-q}.  Geometric optics approximation is sufficient to describe it in   near-Earth environments. In this framework   a careful analysis of the local standard polarization directions allows one to obtain transparent expressions for polarization rotation \cite{tg-pol}. In particular, it clarifies the statement about the absence of polarization rotation in Schwarzschild space-time.  For an open trajectory this is true only for a particular choice of   local polarization directions (that we call the Newton gauge and review below). For a closed (in phase space) trajectory the net rotation disappears in any gauge. Given that the leading order corrections due to gravity result from the leading terms in the post-Newtonian expansion of the Schwarzschild metric \cite{postN,will},  this result seems to indicate that the Faraday rotation can be ignored. However, it was pointed out in Ref.~\cite{gf-yes} that in realistic situations  the closed spatial trajectories are   not closed in the phase space, as well as that enforcing the Newton gauge is in general impractical. Hence, even if the gravitationally induced polarization rotation in the near-Earth environment  is essentially a pure gauge effect,  it cannot be simply dismissed.

We provide a simple estimation of this emitter- and observer-dependent phase and give its explicit form in several settings. The   phase errors due to  SR can be mitigated, and in the limit of sharply defined frequencies, completely removed by a simple encoding procedure \cite{bt:05}. The same techniques are applicable to counter the Faraday rotation.

The rest of this paper is organized as follows. In Sec.~\ref{wigner} we review the SR effects. Polarization rotation in general stationary space-times is described in Sec.~\ref{fara}. In Sec.~\ref{phases} we evaluate the effects in communication with Earth-orbiting satellites.  In Sec.~\ref{errors} we present a simple error-correcting scheme, discuss  the role of the relativistic errors (as well as errors due to the reference  frame misalignment), in quantum key distribution (QKD), and discuss the feasibility of the error-correcting schemes.  This is a fully reference frame independent (RFI) scheme that may be useful not only for countering tiny relativistic effects in communications with the LEO satellites,  but also as the major source of the phase errors --- frame misalignment.

 We work with $c=\hbar=G=1$. The  constants $G$ and $c$ are restored in a small number of expressions where their presence is helpful. The space-time metric $\sg_{\mu\nu}$ has a signature $-+++$. The four-vectors are distinguished by the sans font, $\ak$, $k^\mu=(\ak)^\mu$. The three-dimensional spatial metric is denoted as $\gamma_{mn}$, and three-dimensional vectors are set in boldface, $\bk$, or are referred to by their explicit coordinate form, $k^m$. The inner product in the metric $\gamma$ is denoted as $\bk\!\cdot\!\bof$, and the unit vectors in this metric are distinguished by the caret, $\hbk$, $\hbk\cdot\hbk\equiv1$. Post-Newtonian calculations employ a fiducial Euclidean space. Euclidean vectors are distinguished by arrows, $\vk$. Components of the two types of vectors may coincide,  $(\bv)^m=(\vb)^m$, but $\vb\!\cdot\!\vk=\sum_{k=1}^3 v^m k^m$. Accordingly, the coordinate distance is the Euclidean length of the radius vector, $r\equiv\sqrt{\vx\!\cdot\!\vx}$.

\section{Wigner rotation} \label{wigner}

Quantum states of a photon with a definite four-momentum $\ak=(|\bk|,\bk)$ can be represented either as Hilbert space vectors or as complex polarization vectors in the usual three-dimensional space,
\be
|\Psi_\ak\9=f_+|\ak,+\9+f_-|\ak,-\9 \quad \Leftrightarrow \quad \hbf_\bk=f_+{\hbb}_\hbk^++f_-{\hbb}_\hbk^-,
\ee
with the transversal vectors $\hbb^\pm_\hbk$, $\bk\cdot\hbb^\pm_\hbk=0$. The correspondence is rooted in the relationship between   finite-dimensional and unitary representations of the Poincar\'{e} group \cite{wig,wkt}.

Unitary  transformations $U$ of states of elementary particles are obtained  via the induced representation of the Poincar\'{e} group. Basis states that correspond to   arbitrary momenta $\bk$ are defined with the help of the standard Lorentz transformation $L(\ak)$  that takes the four-momentum from the standard value $\ak_S$ to $\ak$.  The direction $\hbk$ is determined by the spherical angles
$(\theta,\phi)$. For massless particles $k_S^\mu=(1,0,0,1)$ and
 \be
 L(\ak)= R(\hbk) B_z(\xi_{|\bk|}),
 \ee
 where $R(\hbk)=R_z(\phi)R_y(\theta)$ rotates the $z$ axis into the direction $\hbk$ by performing rotations around the $y$ and $x$ axes by the angles $\theta$ and $\phi$, respectively. These rotations follow the boost $B_z(\xi_{|\bk|})$ along the $z$ axis that brings the magnitude of the momentum to $|\bk|$ (the rapidity $\xi_{|\bk|}\geqslant0$ is determined by $\sinh\xi+\cosh\xi=|\bk|$).

 The states of an arbitrary momentum are defined via
 \be
 |\ak,\pm\9\defeq U\big(L(\ak)\big)|\ak_S,\pm\9,
 \ee
 while the standard right and left-circular polarization vectors are defined via
 \be
 \hbb_\hbk^\pm\defeq R(\hbk)(\hat{\mathbf{x}}\pm i \hat{\mathbf{y}})/ \sqrt{2},
 \ee
 while the linear polarization vectors are $\hbb_1\defeq R(\hbk)\hbx$ and $\hbb_2\defeq R(\hbk)\hby$, respectively. Alternatively, these vectors can be obtained as
 \be
 \hbb_2=\frac{\hat{\mathbf{z}} \times \hat{\mathbf{k}}}{|\hbz\times\hbk|}, \qquad \hbb_1=\hbb_2\times\hbk. \label{b12k}
 \ee
  The explicit form of the polarization  four-vector  $\af_\ak$,  $\af_\ak\cdot \ak=0$ depends on the gauge
 \cite{pt:04,fara-q,hks:85}.

 The Wigner stability subgroup consists of all proper Lorentz transformations that leave the standard four-momentum $\ak_S$ invariant:
\be
W(\Lambda, \ak) =L^{-1}(\Lambda \ak)\Lambda L(\ak). \label{wig0}
\ee
For massless particles it is decomposed as
\be
W=S R_z(\varpi), \label{wdec}
\ee
where $S$ is a translation in the $xy$ plane and $R_z(\varpi)$ is the rotation around the $z$ axis. As the translations $S$ do not correspond to physical degrees of freedom,
the state transforms according to
\be
U(\Lambda) |\ak,\pm\9=e^{\pm i \varpi}|\Lambda \ak,\pm\9. \label{phase1}
\ee

There are no generic explicit  expressions for $\varpi$. Their evaluation is  not considerably simpler if $\Lambda=\eR$, where $\eR$ is a rotation (as there is no risk of confusion we use the same designation for the four-dimensional matrices of spatial dimensions and for their $3\times 3$ blocks). However, as the transformation law of $\hbf$ can be obtained from the three-dimensional form of the Lorentz transformations of the transversal electromagnetic wave, in this case \cite{pt:04,lpt-1}
\be
U(\eR)|\Psi_\ak\9  \Leftrightarrow \eR \hbf_\bk.
\ee
Moreover, an arbitrary  rotation around the  direction $\hbb_2$, $R_{\hbb_2}(\alpha)$,
does not introduce a phase $\varpi$ \cite{tg-pol}. This provides the motivation for introduction of the so-called Newton gauge that we review below.

\section{Faraday rotation}\label{fara}

The equations of geometric optics are obtained  by performing the short-wave asymptotic expansion of the wave equation in the Lorentz gauge. The  four-vector potential is represented as
$A^\mu(x)=a^\mu e^{i\psi}$. {The} wave vector, $k_\mu:=\pad_\mu\psi$, is by definition  normal to hypersurfaces of constant phase $\psi$; {in addition} it is null, $k^{\mu}k_{\mu}=0$.  Hence hypersurfaces of constant $\psi$ are null and their normals are also tangent vectors to the null geodesics $x^\mu(\sigma)$ that lie in them \cite{mtw}:
\be
\frac{dx^\mu}{d\sigma}=k^\mu, \qquad k^{\mu} \nabla_{\mu} k^{\nu} = 0. \label{mom1}
\ee
Here $\nabla_{\mu}$ is a covariant derivative compatible with the background metric $\sg$ and $\sigma$ is the affine parameter along a geodesic. 

The eikonal equation, which is a restatement of the null condition, is given by
\be
\sg^{\mu\nu}\frac{\pad\psi}{\pad x^\mu}\frac{\pad\psi}{\pad x^\nu}=0, \label{HJequation}
\ee
{which} is the Hamilton-Jacobi equation for a free massless particle on a given background  {space-time}.

 The polarization four-vector is defined as  $f^\mu:=a^\mu/\sqrt{a^\mu a_\mu^*}$. It is transversal to the  {null} geodesic  {generated by $k^{\mu}$ and is parallel propagated along it}:
\be
f^\mu k_\mu=0,\qquad k^\mu\nabla_\mu f^\nu=0. \label{pol1}
\ee
 {Thus} we treat photons as massless point particles that move on the rays prescribed by the geometric optics.

Stationary space-times allow a convenient three-dimensional representation of the evolution of the polarization vector \cite{fl:82,tg-pol}. Static observers follow the congruence of timelike Killing vectors that  define a projection from the  space-time manifold $\cal{M}$ onto a three-dimensional space $\Sigma_3$, $\pi:\mathcal{M}\rightarrow \Sigma_3$. In practice, this projection is performed by dropping the timelike coordinate of an event, and  vectors are projected via a push-forward map: $\pi_*\ak=\bk$.

The metric $\sg_{\mu \nu}$ on $\mathcal{M}$ can be written in terms of a three-dimensional scalar $h$, a vector $\bg$ with components $\sg_m$, and a three-dimensional metric $\gamma_{mn}$ on $\Sigma_3$ as
\be
ds^2=-h \left( dx^0 - \sg_m dx^m \right)^2+dl^2,
\ee
where $h:=-\sg_{00}$, $\sg_{m}:=-\sg_{0m}/\sg_{00}$, and the three-dimensional distance is  $dl^2:=\gamma_{mn}dx^mdx^n$, with the three-dimensional metric
\be
\gamma_{mn}=\sg_{mn}-\frac{\sg_{0m}\sg_{0n}}{\sg_{00}},
\ee
whose inverse is $\sg^{mn}$.
The three-dimensional $\gamma_{mn}$-compatible covariant derivative is denoted as  $D_m$.

Using the relationships between the three- and four-dimensional covariant derivatives, the propagation equations \eqref{mom1} and \eqref{pol1} in a stationary spacetime result in the following three-dimensional expressions  \cite{fl:82,tg-pol}:
\begin{align}
\frac{D\hbk}{d\sigma} &= \Omegab \times \hbk, \label{3dmrot}  \\
\frac{D\hbf}{d\sigma} &= \Omegab\times\hbf, \label{3dprot}
\end{align}
where $\sigma$ is an affine parameter along the curve with tangent vector $\mathbf{k} = d\mathbf{x}/d\sigma$. From Eqs. \eqref{3dmrot} and \eqref{3dprot} we see that both the propagation direction and polarization are rigidly rotated, with an angular velocity given by  \cite{tg-pol}
\be
\Omegab=2\omegab-(\omegab\!\cdot\!\hbk)\hbk-\bE_g\times\bk, \label{Omegab}
\ee
with the vector $\left(h, \bg \right)^T$ playing the role of a vector potential for gravitoelectric and gravitomagnetic field \cite{mtw, inertia}:
\be
\bE_g=-\frac{\nabla h}{2 h}, \qquad  \omegab=-\half k_0 \nabla \times \mathbf{g}. \label{valom}
\ee

In flat spacetime this basis is uniquely fixed by Wigner's little group construction.
However, on a general curved background the Wigner construction must be performed at every point, i.e.,  absence of a global reference direction ensures that the standard polarization triad $(\hbb_1, \hbb_2,\hbk)$ is different at every location. Given such choice  the net polarization rotation can be found   by starting with the initial polarization
$\af_{\mathrm{in}}(x_\mathrm{in})=\arb_1(x_\mathrm{in})$, parallel propagating it according to Eq.~\eqref{pol1}, and reading off the
angle from the decomposition of $f_{\mathrm{fin}}(x_\mathrm{fin})$,
\be
\af=\cos\chi \arb_1+\sin\chi \arb_2 \label{polangle1}.
\ee

Evaluation of the polarization rotation is much simpler in stationary spacetimes.  By setting $\hbf=\hbb_1$ at the starting point of the
trajectory, we have $\sin\chi=\hbf\cdot\hbb_2$,
 so
\begin{align}
\frac{d\chi}{d\sigma}&=\frac{1}{\hbf\!\cdot\!\hbb_1}\frac{D(\hbf\!\cdot\!\hbb_2)}{d\sigma} \nonumber \\
&=\omegab\!\cdot\!\hbf+\frac{1}{\hbf\!\cdot\!\hbb_1}\hbf\!\cdot\!\frac{D\hbb_2}{d\sigma}. \label{polangle}
\end{align}

In the Schwarzschild spacetime $\omegab\equiv 0$ and polarization rotation is a pure gauge effect. The phase remains zero if the standard directions are set with the help of the local free fall acceleration $\mathbf{w}$ of a stationary observer. At each point in the spacetime we choose the direction of the standard reference momentum, or equivalently the $z$-axis of our standard polarization triad, to be $\hzeb\defeq\mathbf{w}$.  For a photon with momentum $\mathbf{k}$ we choose the linear polarization vector $\hat{\mathbf{b}}_2$ to  point in the direction $\hzeb \times \hat{\mathbf{k}}$, and finally we choose $\hat{\mathbf{b}}_1\defeq \hbb_2\times\hbk$ such that it completes the orthonormal triad $( \hat{\mathbf{b}}_1, \hat{\mathbf{b}}_2, \hat{\mathbf{k}})$. This construction is known as the Newton gauge \cite{tg-pol}. With this convention
$\Omegab=-\bE_g\times\bk\equiv\Omega \hbb_2$ and thus $\chi\equiv 0$ along the trajectory.

However, such choice of standard polarization direction is practically unfeasible. We will see the consequences of setting the $z$-axis with the help of a guide star in the next section.

 \begin{figure}[htbp]
\includegraphics[width=0.38\textwidth]{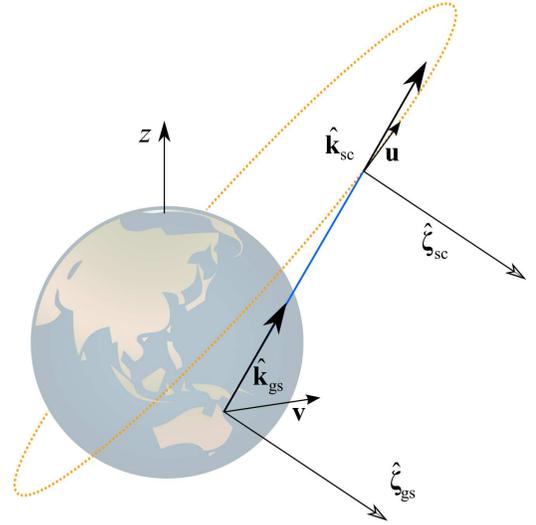}
\caption{ \label{schema} Scheme of the communication round between the ground station (GS) and the spacecraft (SC). The light ray is highlighted in blue.
 All the vectors are indicated in the Earth-centered inertial frame. In the flat space-time approximation the
 unit tangent vectors to the ray $\hbk_{\mathrm{gs}}$ and $\hbk_{\mathrm{sc}}$ at the GS and the SC, as well as the direction to an infinitely distant guide star, $\hzeb_{\mathrm{gs}}$ and $\hzeb_{\mathrm{sc}}$, respectively, coincide.  Velocity of the GS at the emission of the signal is $\bv$ and velocity of the SC at the moment of detection is $\bu$.  }
\end{figure}

\section{Relativistic phases in the near-Earth quantum communications}\label{phases}

\subsection{Setting}
As a typical scenario we consider  one round of communications between the ground station (GS) and a LEO spacecraft (SC).
The problem is most conveniently analyzed in the geocentric (Earth-centered inertial) system, the origin of which coincides with the center of the Earth at the moment of emission. We direct the $z$ axis of this system along the Earth's angular momentum. Then the three-dimensional velocity $\bv$ of the GS in this frame lies in the $xy$ plane, while the initial propagation direction vector $\hbk$ and the velocity of the SC at the time of detection $\bu$, both expressed in the global frame, are arbitrary.  Fig.~\ref{schema} represents this scenario.

The parametrized post-Newtonian approximation \cite{postN} is  a systematic  method  for obtaining corrections to the Newtonian motions of  a system of slowly moving bodies bound together by  weak gravitational forces. For a metric theory in question the corrections are organized by the powers of in $GM/rc^2$ and $v/c$. Typical velocities and gravitational potentials are related by $GM/r \sim v^2$. By estimating $\epsilon^2\defeq GM/R_\oplus c^2\approx 10^{-5} $, where $R_\oplus$ is the Earth's radius we find that typical velocities are of the order $\epsilon$,  and the upper bound on the gravitational potential is of the order $\epsilon^2$. The order of expansion is conveniently labeled by the parameter $\epsilon$, the formal powers of which accompany the corresponding expressions and that is taken to unity at the end of the calculation.

\begin{figure*}[t]
    \centering
        \includegraphics[width = 0.47\textwidth]{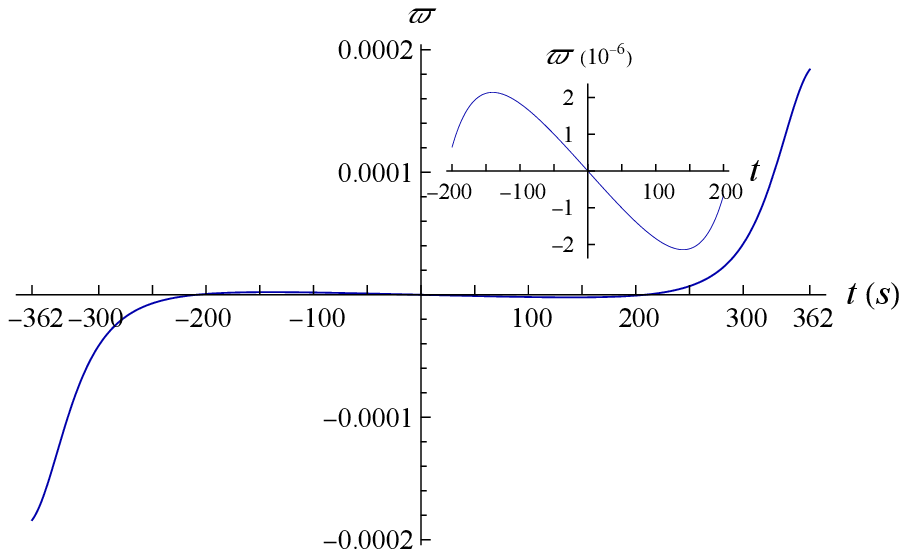}
    \includegraphics[width = 0.47\textwidth]{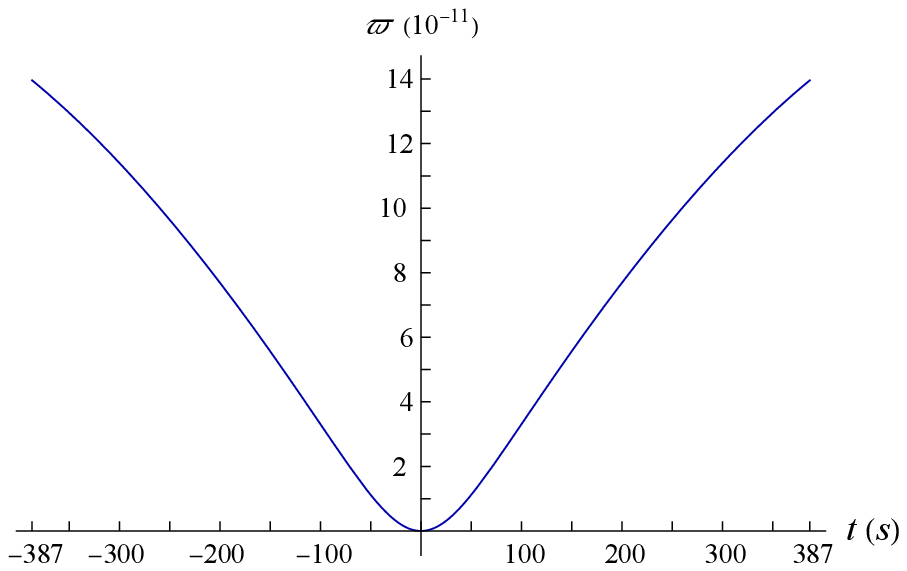}
      \caption{ \label{wigphase} { Wigner phase in one round of the GS to SC communication as a function of the passage time. The GS is located on the equator ($r_\gs=R_\oplus=6.38\times 10^6$m) and the SC has a circular orbit ($r_{\mathrm{sc}}=7.00\times 10^6$m). The time origin corresponds to the SC in the zenith. {\it (left)} The SC moves on a constant celestial meridian and crosses the equator at $t=0$. The inset shows the Wigner phase when the SC is close to the zenith. {\it (right)} The SC rotates in the equatorial plane in the same sense as the Earth. For the coplanar motion the first-order effect vanishes, and the Wigner phase is of the same order of magnitude as the Faraday phase of Sec.~\ref{Fphase}. }}
 \end{figure*}

Hence the leading  corrections that take into account the effects of both SR and GR include the terms of up to the order of $\epsilon^2$.   In this approximation the gravitational field of the Earth is spherically symmetric, and thus it follows from Eq.~\eqref{polangle} that the perceived polarization rotation results only due to the reference frame choices by the communicating parties.
Moreover, at this level of precision the effects of SR and GR can be treated separately, as the leading  mixed term is of the order $\epsilon^3$.

\subsection{SR effects}

It is easy to see that the SR effects dominate, resulting in $\varpi\sim \epsilon$. It also should be noted that the freedom in choosing the relative orientation of the reference frames allows one to eliminate the Wigner phase for a fixed relative velocity $\bv_{12}$ between the two frames 1 and 2,  and any four-momentum vector $\ak_1$ (expressed, say, in   frame 1). This is similar to the two-frame invariant definition of spin of massive particles \cite{cw:03}. Indeed, any proper Lorentz transformation that relates these two frames has a form \cite{wkt}
\be
\Lambda=R(\hbv_{12})B_z(\xi_{v_{12}})R(\alpha,\beta,\gamma)^{-1},
\ee
where $v_{12}=\tanh\xi_{v_{12}}$ and $R(\alpha,\beta,\gamma)$ is an arbitrary (four-dimensional matrix of) rotation with the Euler angles $\alpha$, $\beta$, $\gamma$. By choosing $R(\alpha,\beta,\gamma)=R(\hbk)$
we have in Eq.~\eqref{wig0} $\Lambda L(\ak)=L(\Lambda\ak)$, ensuring absence of the Wigner phase.

We use the setting  of Fig.~\ref{schema} and the  conventions of Sec.~\ref{wigner} (so the standard propagation direction is along the $z$ axis of the Earth-centered inertial frame). As we want to focus on the effects of relative motion we assume that the global frame and the GS and SC frames are related by pure boosts, $B(\bv)$ and $B(\bu)$, respectively.
Then the momenta in GS and SC frames are related by
\be
k_\SC=B(-\bu)B(-\bv)k_\GS, \label{lor12}
\ee
and thus the Wigner rotation is obtained by extracting the angle $\varpi$ from the decomposition of the Lorentz transformation $\Lambda=B(-\bu)B(-\bv)$ via Eq.~\eqref{wdec}.
We take the velocities of the GS and SC in the global frame as
\begin{align}
&\bv=(v\cos\alpha,v\sin\alpha,0), \\ &\bu=u(\sin\vartheta\cos\varphi,\sin\vartheta\sin\varphi,\cos\vartheta),
\end{align}
respectively, and the propagation direction
\be
\hbk=(\sin\theta\cos\phi,\sin\theta\sin\phi,\cos\theta).
\ee
The first-order contribution to the Wigner phase is
\be
\varpi= u\cot \theta \sin\vartheta \sin (\phi-\varphi)+ v \cot \theta \sin (\phi-\alpha)    +\co(\epsilon^2).
\ee
For the SC moving at the latitude of the GS coplanar the first-order effect is absent.

Fig.~\ref{wigphase} illustrates the Wigner phase accrued between the GS and the SC  as a function of the SC passage time.  {As our goal is to illustrate the relativistic effects in their simplest settings we locate the ground station on the equator and assume an exactly circular trajectory for the spacecraft.   Except for the coplanar motion where the first-order Doppler effect cancels, the induced phase is of the order of $v/c$, where $v$ is the relative velocity between the GS and SC.

\subsection{GR effects} \label{Fphase}


Electromagnetic radiation and massless particles are not affected by Newtonian gravity. The leading post-Newtonian contributions (corrections to trajectories, time differences, and phases) are of order $\epsilon^2$ in the parametrized post-Newtonian (PPN) expansion \cite{postN,will,mtw,inertia}; to take into account gravitomagnetic effects we need contribution up to $\epsilon^3$.

The post-Newtonian expansion of the metric near a single slowly rotating quasirigid gravitating body, assuming that the underlying theory of gravity is GR \cite{postN,will,mtw}, up to terms of order $\epsilon^3$ is given by
\be
ds^2=-V^2(r)c^2dt^2+\vec{R}\!\cdot\!d\vx\, cdt+W^2(r)d\vx\!\cdot\!d\vx, \label{farfieldkerr}
\ee
where
\be
V(r)=1-\epsilon^2\frac{U}{c^2}, \qquad W(r)=1+\epsilon^2\gamma\frac{U}{c^2}.
\ee
The Newtonian gravitational potential $-U=-GMQ(r,\theta)/r\simeq-GM/r$  depends on the mass $M$ and higher multipoles \cite{will,inertia}, and
\be
\vec{R}=-\epsilon^3 4\frac{G}{c^3}\frac{\vJ\times\vx}{r^3},
\ee
where $\vec{J}$ is the angular momentum {of the rotating body}. Hence we see that the gauge-invariant polarization rotation is absent in the leading order of the post-Newtonian expansion and the Faraday phase at
the order $\epsilon^2$ is a reference-frame effect. As we are interested only in the leading contribution to the phase we set $Q=1$, and as it limits the precision at the level of $10^{-3}$ \cite{inertia},  $\gamma=1$. In the remainder of the section we revert to the units $G=c=1$.

To obtain the leading contributions to the phase {and polarization rotation}, the photon trajectories only need to be expanded up to $\epsilon^2$,
\be
\vx(t)=:\vx_{(0)}(t)+\epsilon^2\vx_{(2)}(t),
\ee
where the zeroth-order Newtonian trajectory is determined by the initial position $\vx(t_0)\equiv \vx_0$ and the initial direction $\vn$, $\vn\cdot\vn=1$. The leading-order corrections are decomposed into the
tangential and transversal components,
\be
\vx_{(2)}(t)=\vn x_\parallel(t)+\vx_\perp(t),
\ee
with $\vx_\perp(t)\cdot \vn=0$ and the initial conditions $\vx_{(2)}(t_0)=0$.
The initial tangent vector satisfies \cite{will}
\be
\vk(t_0)\defeq\left.\frac{d\vec{x}}{dt}\right|_{t_0}=(1-2\epsilon^2U(\vec{x}_0))\vn, \qquad \!\vn\!\cdot\vn=1. \label{kGS}
\ee
Since $\gamma_{mn}k^mk^n=W^2V^4$,   then with a slight abuse of notation in leading post-Newtonian order we have
\be
 \hbk_0=W_0\vk_0.
\ee

The post-Newtonian corrections are obtained either from the evolution of the unit Euclidean vector $\vv$ or from the equations
\begin{align}
&\frac{d^2 \vx_\perp}{dt^2}=2\big (\vec\nabla U-\vn(\vn\!\cdot\!\vec\nabla U)\big), \\
&\frac{dx_\parallel}{dt} =-2 {U}. \label{pardel}
\end{align}

\begin{figure*}[tb]
    \centering
        \includegraphics[width=0.33\textwidth]{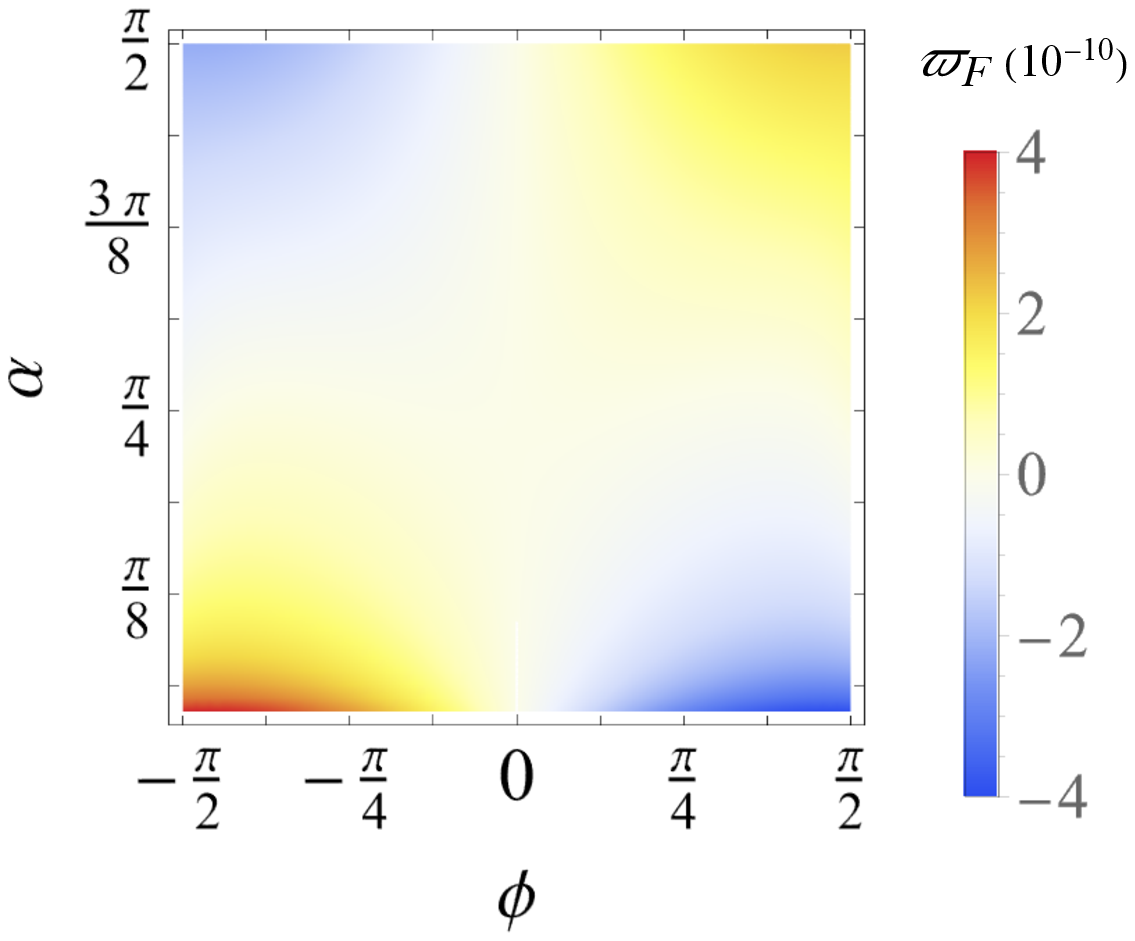}
         \includegraphics[width=0.33\textwidth]{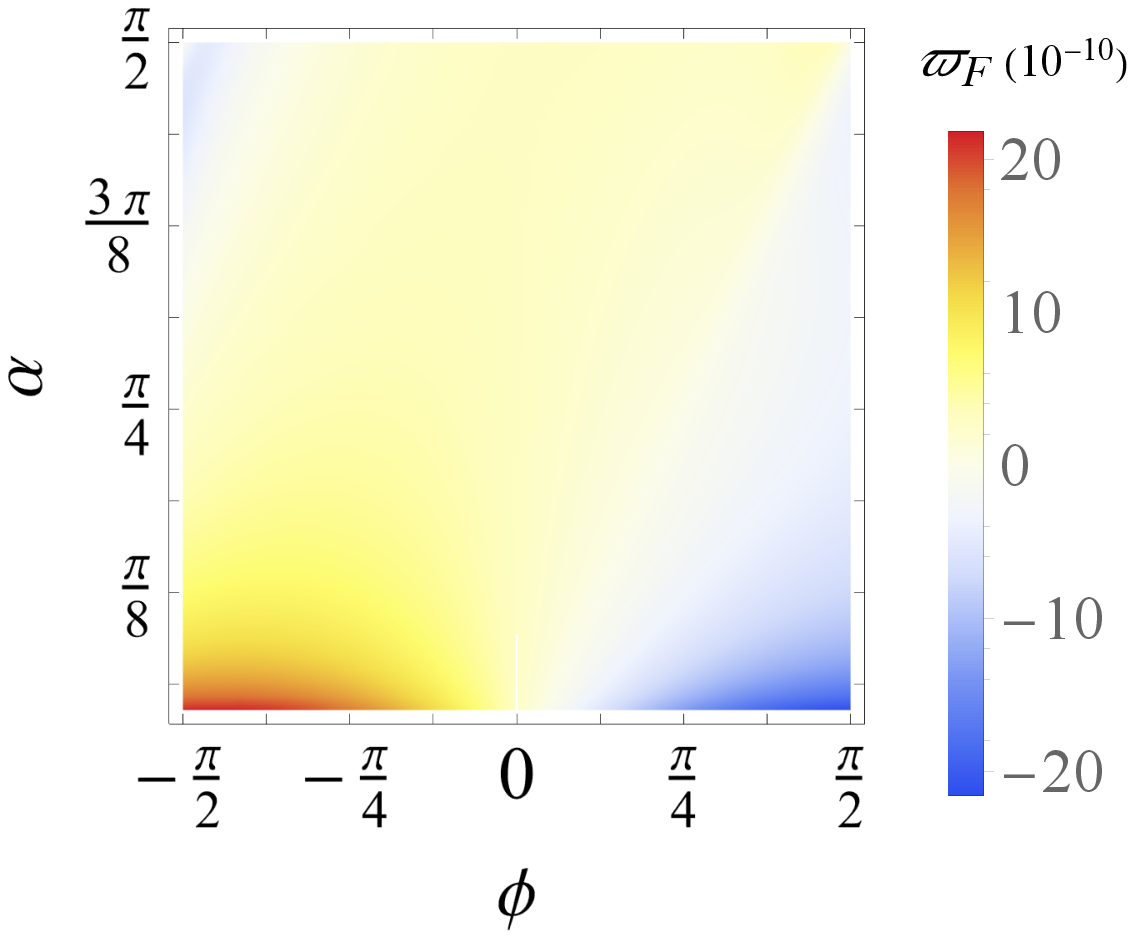}
          \includegraphics[width=0.33\textwidth]{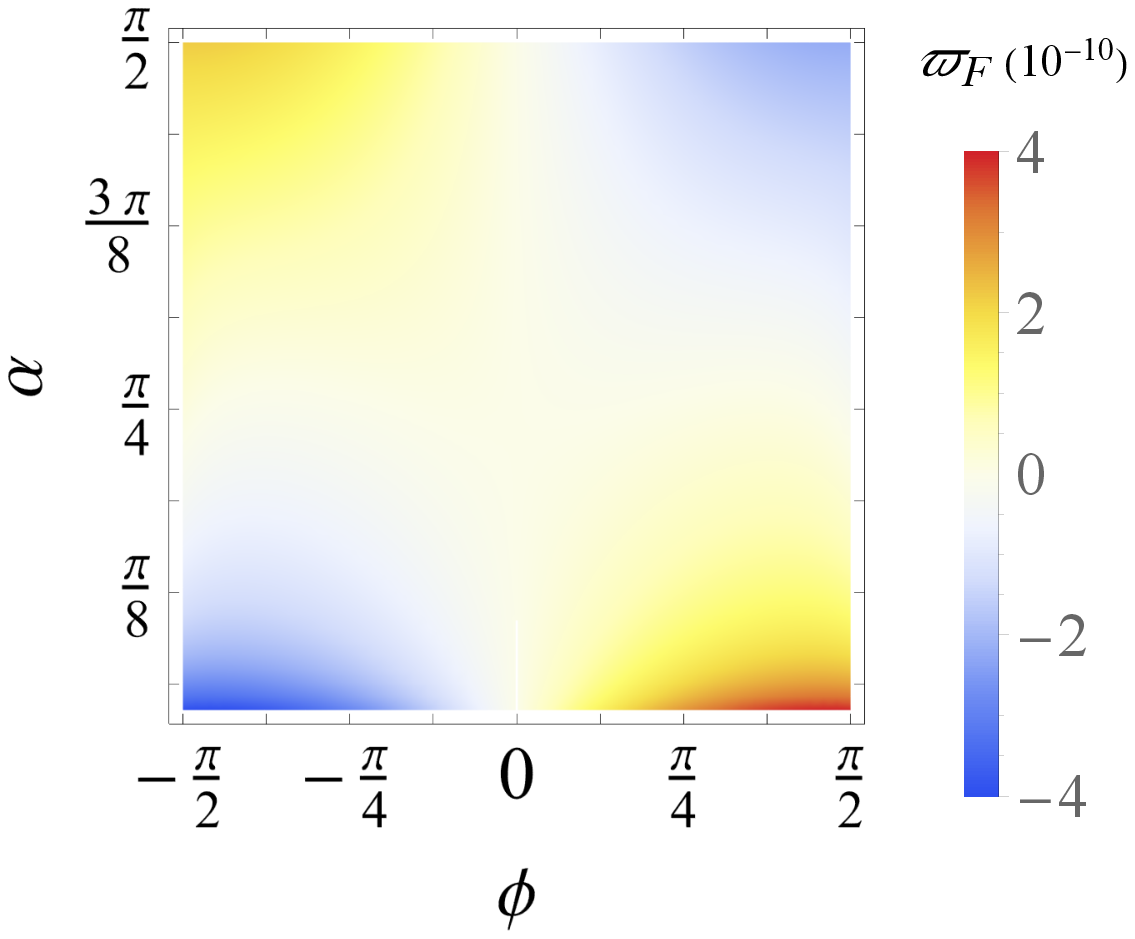}
      \caption{ \label{Faraphase} The Faraday phase depends on the choice of the reference direction $\vec{\zeta}(\alpha,\beta)$ and the initial propagation direction $\vn(\phi)$. The angles are defined in Eqs.~\eqref{vn} and \eqref{vl}. {\it (left)} $\beta=0$. {\it (center)} $\beta=7\pi/16$. {\it (right)} $\beta=\pi$. }
    \label{fig:mink}
\end{figure*}

With the origin at the center of the Earth the integration results in the   tangent vector
\be
\vk=\vn-\epsilon^2\frac{2M\vn}{r(t)}-\epsilon^2\frac{2M\vd}{d^2}\left(\frac{\vx(t)\cdot\vn}{r(t)}-\frac{\vx_0\cdot\vn}{r_0}\right), \label{k2}
\ee
    where it is enough to take the Newtonian equation of the trajectory to obtain the leading post-Newtonian correction,
\be
\vx(t)\cdot \vn=\vx_0\cdot\vn+t-t_0+\co(\epsilon^2),
\ee
and
\be
\vd\defeq \vx_{0}-(\vx_{0}\cdot\vn)\vn,
\ee
is the vector joining the center of the Earth and the point of the closest approach of the unperturbed ray.

The leading-order post-Newtonian metric is spherically symmetric. The rest of the calculations are considerably simplified if we note that the initial polarization that is perpendicular to the propagation plane remains perpendicular to it, which is an immediate consequence of Eqs.~\eqref{3dprot}--\eqref{valom}.  Therefore we select the reference frame differently from that of Fig.~\ref{schema}. As the effects of SR contribute at least one additional factor of $\epsilon$, here we focus only on the gravitational effects and treat them separately from the effects of rotation and relative motion. Hence  take as $z=0$ the plane where the ray from the GS to the SC lies, set their velocities to zero, and consider the (constant Euclidean) polarization vector
\be
\vf=(0,0,1)=\mathrm{const}.
\ee
The gauge-dependent Faraday phase results from the changes in the definitions of the standard polarization directions along the trajectory. In turn, these follow from the changes in the standard polarizations. At each point they are defined via Eqs.~\eqref{b12k} where the unit vector $\hzeb$ pointing to a distant reference star takes the role of  $\hbz$ and the
local unit tangent $\hbk$.

The local propagation direction at the GS and the SC is conveniently represented:
\begin{align}
\vk_i&=\vn\left(1-\epsilon^2\frac{2M}{r_i}\right)-\epsilon^2\frac{2M\vd}{d^2}\left(\frac{\vx_i \cdot\vn}{r_i}-\frac{\vx_\gs\cdot\vn}{r_\gs}\right) \nonumber\\
&\eqdef \vn+\epsilon^2\vkap_i, \qquad i=\mathrm{gs,sc}.
\end{align}
The reference directions $\hzeb_i$, $i=\mathrm{gs, sc}$ are obtained from the  tangents to the rays from the fixed guide star that arrive to the GS and SC, respectively, and $r_\gs=R_\oplus$.
 We assume that the star is infinitely far, so in the Newtonian limit $\hzeb_i^m\to l^m$, where $-\vl$ is the flat space direction from the infinitely distant star to the observers, $\vl\cdot\vl=1$. Approximating  differences in the local directions as arising solely from the gravitational field of the Earth, we have
\be
\vec\zeta_i=\vl\left(1-\epsilon^2\frac{2M}{r_i}\right)-\epsilon^2\frac{2M\vd_i}{d^2_i}\left(\frac{\vx_i\! \cdot\!\vl}{r_i}-1\right)\eqdef \vl+\epsilon^2\vec\varsigma_i.
\ee
Then  the standard polarization vectors at the GS and the SC are
\be
\hbb_{2i}=\frac{\vec{\zeta}_i\times\vk}{\big|\vec{\zeta}_i\times\vk\big|}\eqdef\ve_2+\epsilon^2\vec\beta_{2i}, \qquad \hbb_{1i}=\hbb_{2i}\times\hbk\eqdef\ve_1+\epsilon^2\vec\beta_{1i}.
\ee
In the Newtonian limit,
\be
\hbb_{2i}\rightarrow \vec{e}_2\defeq\frac{\vl\times\vn}{\big|\vl\times\vn\big|},
\ee
and to ensure  $|\hbb_2|=1$ the post-Newtonian terms satisfy $\ve_2\cdot\vec\beta_{2i}+M/r_i=0$,
with   analogous expressions  for $\hbb_{1i}$. Explicitly,
\begin{align}
&\ve_2=\frac{\vl\times\vn}{\sqrt{1-(\vl\!\cdot\!\vn)^2}},\\
& \vec{\beta}_{2i}=\frac{\vl\times\vn\,(\vl\!\cdot\!\vn)(\vn\!\cdot\!\vec{\varsigma_i}+\vl\!\cdot\!\vkap_i)+\vl\times\vkap_i-\vn\times\vec{\varsigma_i}}{\sqrt{1-(\vl\!\cdot\!\vn)^2}}.
\end{align}


 As a result, the leading order Faraday phase that is accrued between the  GS and the SC is obtained from
\begin{align}
\sin(\chi+\Delta\chi)- \sin\chi=\hbf_\sc\!\cdot\!\hbb_{2\sc}-\hbf_\gs\!\cdot\!\hbb_{2\gs} \nonumber \\
=\Delta\chi\cos\chi=\Delta\chi\vf\!\cdot\ve_1+\co(\epsilon^3).
\end{align}
Substituting  the  explicit expressions into the above equation results in
\be
\Delta\chi=\frac{e_2^zM}{e_1^z}\left(\frac{1}{r_\gs}-\frac{1}{r_\sc}\right)+\frac{\beta_{2\sc}^z-\beta_{2\gs}^z}{e_1^z}. \label{phase2}
\ee
where we assume that $e_1^z\gg \epsilon^2$, which is the validity condition for this expression for the Faraday rotation.

If we choose the $x$ axis to pass through the GS, then
\be
\vn=(\cos\phi,\sin\phi,0), \qquad -\half\pi<\phi<\half\pi, \label{vn}
\ee
and
\be
\vl=(\cos\alpha,\sin\alpha\sin\beta,\sin\alpha\cos\beta),\label{vl}
\ee
where the altitude of the guide star is $\half\pi-\alpha$, $0\leqslant \alpha\leqslant \half \pi$. Some useful auxiliary expressions  are given in the Appendix.

When the reference direction $\hzeb$ and the propagation direction $\hbk$ are collinear, the  standard polarization directions are undefined. If $\hzeb$ lies in the plane determined by GS, SC, and centre of the Earth, then the Faraday phase is zero. Moreover, the post-Newtonian expansion of Eq.~\eqref{phase2} fails
  in the $\epsilon^2$ vicinity of $\alpha=0$ or $\beta =\half \pi,\tfrac{3}{2}\pi$.

While for a general configuration an  expression for the Faraday phase is rather cumbersome, some special cases are quite simple. For $|\tan\beta|\ll\epsilon^{-2}$ and the SC at the zenith ($\phi=0$) the Faraday phase is
\be
\varpi_{\mathrm F}=\frac{GM}{c^2}\frac{r_\sc-r_\gs}{r_\sc r_\gs}\tan\beta,
\ee
while for the guide star close to  the zenith the resulting phase  is
\be
\varpi_{\mathrm F}=-\frac{GM}{c^2}\frac{r_\sc-r_\gs}{r_\sc r_\gs}\frac{1}{\cos\beta}\left(\frac{\sin\phi}{\alpha}-\cos\phi\sin\beta+\co(\alpha)\right).
\ee
for $\epsilon^2\ll\alpha\ll1$.

 Fig.~\ref{Faraphase} shows the leading-order (gauge-dependent) Faraday phase for different reference and propagation directions.  As discussed above, to isolate the gravitational contribution to the phase
it is enough to consider a fictitious scenario of the GS and SC being in fixed positions in the global reference frame. Unless the first-order Doppler effect cancels, the Faraday phase is $\epsilon$ times smaller than the Wigner phase.}

\section{Errors and their mitigation} \label{errors}
Regardless of the basis that is used to encode polarization qubits the relativistic phases lead to errors in distinguishing the signals. For small phases the fidelity loss, $1-F$, $F\defeq|\6\Psi_\sc|\Psi_\gs\9|$, scales as $\varpi^2$.
 The SR and GR effects appear jointly and can be countered using the same techniques. As the phases are opposite for the opposite helicities, for the states with sharply defined momenta
the phase errors can be easily countered \cite{bt:05}.

Consider
two entangled well-separated and therefore distinguishable
wave packets, with the same momentum profile centered on
$\ak$. For example, the two-photon states
\be
|\Psi^\pm_\ak\9=\frac{1}{\sqrt{2}}\big(|\ak,+\9_1|\ak,-\9_2\pm|\ak,-\9_1|\ak,+\9_2\big), \label{encode}
\ee
have a zero helicity and are thus   insensitive to the Wigner and Faraday phase rotations.

The two states $|\Psi^\pm_\ak\9$ are orthogonal to each other and thus perfectly distinguishable. Hence using them one logical qubit can be encoded with two physical
qubits (photons). If multiphoton states can be resolved then the  asymptotically efficient encoding
scheme of using $N$ photons to transmit $N- 2^{-1}\log_2 N$ qubits  can be employed.

The relative importance of the relativistic noise and, therefore, necessity of using this invariant encoding, depend on the specific task and its performance requirements. Consider for definiteness a QKD protocol with discrete variables \cite{l:07,qkd-rmp}, where qubits are realized as polarization states of photons with well-defined momenta. Let the protocol be a version of Eckert's protocol E91, where the entangled  state is prepared onboard of the SC and is sent to the communicating parties on the ground.

 The measured qubit error rates were reported to be 1-3\% \cite{micius-1} for the SC--GS link, raising to 4.5--8.1\% for the entanglement-based QKD that enabled secure communications between two GSs \cite{micius-2}. Weather is the main source of the bit rate variability \cite{micius-1}. In situations like this, where the protocol operates close to the maximal allowed error rate of approximately 11\% \cite{l:07}, the relativistic phase is negligible and should be ignored. Moreover, a nonrelativistic effect --- misalignment of the reference  frames --- is a much more serious problem that potentially may prevent the establishment of a secure key.

As all such misalignments (both relativistic and nonrelativistic) introduce the phase error $|\ak\pm\9\to e^{\pm i\phi}|\ak'\pm\9$, we consider their correction by the RFI protocol of Ref.~\cite{rfi:10}. The protocol is based on the assumption of a shared (logical) $z$-axis by the communicating parties. We denote the logical Pauli operators $\sigma_x\equiv X$, $\sigma_y\equiv Y$ and $\sigma_Z\equiv Z$ of the two parties (Alice and Bob) as $X_{A,B}$, $Y_{A,B}$ and $Z_A=Z_B$, and the logical qubits are realized as $|0\9\defeq|\ak+\9$  and $|1\9\defeq|\ak-\9$, respectively.

Taking the ideal entangled state that is  to be shared between the parties as $|\Psi^-\9$, in the absence of other sources of noise the actual state is
\be
|\Psi^-\9_{AB}=\frac{1}{\sqrt{2}}\big(e^{\phi}|\ak_1+\9_A|\ak_2-\9_B-e^{-i\phi}|\ak_1-\9_A|\ak_2+\9_B\big), \label{state}
\ee
where the relative phase $2\phi$ depends on the momenta of the two photons and the Lorentz transformation between the frames (we discuss weaker gravity effects shortly).

Alice and Bob perform at each step one of the three randomly selected measurements $X$,$Y$, $Z$.   For simplicity we take them as  equiprobable. At each instance when both of them have selected a shared $Z$ Alice and Bob generate a shared bit of the raw key. For the state of Eq~\eqref{state} this shared bit is generated error free (assuming the absence of eavesdropping and other sources of noise), which results in the raw key bit generation rate per communicated pair that is (bounded by) $P_\mathrm{rfi}=1/9$. Four of the eight other possible measurement combinations that involve $X$ and $Y$ tests  are used to generate the correlator
\be
C=\6 X_A X_B\9^2+\6X_A Y_B\9^2+\6Y_A X_B\9^2+\6Y_AY_B\9^2.
\ee
For an arbitrary but constant $\phi$ it is independent of the phase and is used to monitor the eavesdropper's knowledge.

As a matter of principle, the scheme relies on the sufficiently good knowledge of one direction and fails when this is corrupted. This direction is shared by having some physical token. Without active feedback and correction the ideally maintained local directions will lead to the errors of the order of the linear Doppler effect, i.e. $\epsilon\sim 10^{-5}$ with the LEO satellites. However, preservation of this token on its own is unrealistic even in purpose-built systems such as the Gravity Probe B \cite{gpb}. On board of the spacecraft torques proportional to the
angle between the gyro spin vector and the spacecraft roll axis needed to be constantly monitored. A more practical bound is given by the specifications of the CubeSat project, where the systems   aligning the
transmitting telescope with the optical ground station during quantum transmission are expected to operate at the level of percents \cite{qsat}.

As long as two-photon measurements are resource intensive, there is no justification in using the two-to-one scheme of Eq.~\eqref{encode}. On the other hand, once such operations become inexpensive, and increase of the bit rate becomes a priority, the  scheme ~\eqref{encode} will clearly have an advantage, as by being incorporated in the standard E91 protocol it allows one raw  bit per four physical photons, even without using the many-particle encoding.

\section{Summary}
Describing propagation of electromagnetic waves in vacuum  in terms of  rays that follow null geodesics is a very good approximation in the high-frequency regime. Observable deviations
from the geometric optics approximation are expected only in ultrastrong gravitational fields. Within this approximation  the polarization rotation in the Schwarzschild metric, and as a result in the leading post-Newtonian approximation, is a purely gauge effect.

 This phase will be present as a consequence of practical methods of setting up reference frames in the Earth-to-spacecraft communications.    However,   these effects are typically about $10^{-5}$ weaker than the SR effects. If these small errors need to be countered this can be done using the same encoding scheme. Finally, we note that this scheme can serve as a basis of a true RFI protocol that does not assume any shared reference frame information.

\acknowledgments
 DRT thanks Paul Alsing, Fan Jingyun, and Tim Ralph for
useful discussions. PKD thanks Daniel George for advice on
graphical output in Mathematica. This work was supported
by the Grants No. FA2386-17-1-4015 and No. FA2386-20-1-
4016 of the Asian Office of Aerospace Research and Development of the U.S. Air Force Research Laboratory.

\appendix
\onecolumngrid
\section{Explicit expressions}
The zeroth-order standard directions are
\begin{align}
    &\ve_1= \bigg(\frac{\sin \phi \left(\tan \left(\frac{\alpha }{2}\right) \sin \beta \cos \phi+\cos \alpha \left(\tan \left(\frac{\alpha }{2}\right) \sin \beta \cos \phi-\sin \phi\right)\right)}{\sqrt{(\sin \alpha \sin \beta \cos \phi-\cos \alpha \sin \phi)^2+\sin ^2\alpha \cos ^2\beta}},\nonumber\\
    &-\frac{\cos \phi \left(\tan \left(\frac{\alpha }{2}\right) \sin \beta \cos \phi+\cos \alpha \left(\tan \left(\frac{\alpha }{2}\right) \sin \beta \cos \phi-\sin \phi\right)\right)}{\sqrt{(\sin \alpha \sin \beta \cos \phi-\cos \alpha \sin \phi)^2+\sin ^2\alpha \cos ^2\beta}},
    -\frac{(\cos \alpha+1) \tan \left(\frac{\alpha }{2}\right) \cos \beta}{\sqrt{(\sin \alpha \sin \beta \cos \phi-\cos \alpha \sin \phi)^2+\sin ^2\alpha \cos ^2\beta}}\bigg),
\end{align}
and
\begin{align}
    &\ve_2= \bigg(-\frac{(\cos \alpha+1) \tan \left(\frac{\alpha }{2}\right) \cos \beta \sin \phi}{\sqrt{(\sin \alpha \sin \beta \cos \phi-\cos \alpha \sin \phi)^2+\sin ^2\alpha \cos ^2\beta}},\frac{(\cos \alpha+1) \tan \left(\frac{\alpha }{2}\right) \cos \beta \cos \phi}{\sqrt{(\sin \alpha \sin \beta \cos \phi-\cos \alpha \sin \phi)^2+\sin ^2\alpha \cos ^2\beta}},\nonumber\\
    &\frac{\cos \alpha \sin \phi-(\cos \alpha+1) \tan \left(\frac{\alpha }{2}\right) \sin \beta \cos \phi}{\sqrt{(\sin \alpha \sin \beta \cos \phi-\cos \alpha \sin \phi)^2+\sin ^2\alpha \cos ^2\beta}}\bigg).
\end{align}
The first post-Newtonian term of $\vk_\gs$ readily follows from Eq.~\eqref{kGS}, while
the leading correction to the propagation direction at the SC is
\be
    \kappa_{\sc}=
  \frac{2M}{r_\gs r_\sc}\Big(- L+\cos \phi (r_\sc-2 r_\gs),
     \cot \phi \big(\cos \phi (r_\gs-r_\sc)+L\big)-r_\gs \sin \phi, 0 \Big).
\ee
Finally, we quote the deviation of the line of sight to the guide star:
\be
\vec\varsigma_{\GS}=\frac{2M}{r_\gs} \big(2 \cos \alpha-1, (2 \cos \alpha+1) \tan \half\alpha \sin \beta,(2 \cos \alpha+1) \tan \half\alpha \cos \beta\big),
\ee

\twocolumngrid

 \end{document}